\documentclass[twocolumn,showpacs,preprintnumbers,amsmath,amssymb,prl]{revtex4}
\usepackage{amssymb}
\usepackage{amsfonts}
\usepackage{amsthm}
\usepackage{graphicx}
\usepackage{dcolumn}
\usepackage{bm}
\usepackage{color}

\newcommand\mE{{\mathcal E}}
\newcommand\bs{{\boldsymbol s}}
\newcommand\bt{{\boldsymbol t}}
\newcommand\bc{{\boldsymbol c}}

\begin{document}

\title{Quantum Error-Correcting Codes over Mixed Alphabets}

\author{Zhuo Wang$^{1,2}$\footnote{wangzhuo@iphy.ac.cn}}
\author{Sixia Yu$^2$\footnote{yusx@nus.edu.sg}}
\author{Heng Fan$^1$}
\author{C.H. Oh$^2$}
\affiliation{%
$^1$Institute of Physics, Chinese Academy of Sciences, Beijing
100190, China\\
$^2$Centre for Quantum Technologies, National University of
Singapore, 3
Science Drive 2, Singapore 117543}%
\date{\today}


\begin{abstract}
Errors are inevitable during all kinds quantum informational tasks and quantum error-correcting codes (QECCs) are powerful tools to fight various quantum noises. For standard QECCs physical systems have the same number of energy levels. Here we shall propose QECCs over mixed alphabets, i.e., physical systems of different dimensions, and investigate their constructions as well as their quantum Singleton bound. We propose two kinds of constructions: a graphical construction based a graph-theoretical object {\it composite coding clique} and a projection-based construction.
We illustrate our ideas using two alphabets by finding out some 1-error correcting or detecting codes over mixed alphabets, e.g., optimal  $((6,8,3))_{4^52^1}$, $((6,4,3))_{4^42^2}$ and $((5,16,2))_{4^32^2}$ code and suboptimal $((5,9,2))_{3^42^1}$ code. Our methods also shed light to the constructions of standard QECCs, e.g., the construction of the optimal $((6,16,3))_4$ code as well as the optimal $((2n+3,p^{2n+1},2))_{p}$ codes with $p=4k$.
\end{abstract}
\maketitle

Quantum error-correcting code (QECC) \cite{shor,bennett,steane,knill} has been receiving much attention because it plays a vital role in many quantum information tasks such as fault-tolerant quantum computation \cite{ftc}, quantum key distribution \cite{qkd}, entanglement purification \cite{ep} and so on, to fight against inevitable noises. Since its initial discovery, great progress has been made in codes construction, i.e., from stabilizer codes \cite{gottesman0,gottesman1,CRSS0,CRSS1} to nonadditive codes \cite{na1,na2,na3,na4,na5}, from binary case to non-binary case \cite{nb1,nb2,nb3,nb4,nb5}. Among all these constructions, {\it coding clique} \cite{na5,nb5} might be the most powerful one so far. In the past few years, many optimal non-binary codes with particles being of odd \cite{nb5,nb7} or prime power \cite{nb4,nb6,nb8} dimensions have been constructed. However, there still have not an efficient method to find optimal codes of composite dimensions.

All QECCs constructed so far are over single alphabet. However, in the usual case in the laboratory, one may have some recourses being of different dimensions in hand when one want to protect information. Thus it is quite necessary to generalize the standard QECCs to the ones over mixed alphabets. The optimality of quantum codes is decided by the quantum bounds, which are subjected to the tradeoffs among the parameters imposed by the principles of quantum mechanics. In this Letter, we will study how to construct QECCs over mixed alphabets and their quantum Singleton bound \cite{knill}. Also some examples of optimal and suboptimal codes over mixed alphabets are presented.

In what follows we shall illustrate our ideas by the construction of QECCs over 2 alphabets with an obvious generalization to more complicated cases. Here we denote a code over 2 alphabets, i.e., two kinds of physical systems of dimensions $p$ and $q$, by $((n,K,d))_{q^{n_1}p^{n_2}}$, which means that the system has $n_1$ $q$-level particles ($quqit$) and $n_2$ $p$-level particles ($qupit$) with $n_1+n_2=n$. When $n_1=0$ or $n_2=0$, it is reduced to a standard QECC. Depending on whether $p$ and $q$ are coprime or not we propose two different constructions.

In the case of reducible $p$ and $q$, e.g., $q=r\cdot p$ for some integer $r$, a quqit can be regarded as the composite particle of a qupit and qurit. Denote the bit shift and phase shift operators of an $l$-level particle by $X_l=\sum_{j\in\mathbb Z_l}|j+1\rangle\langle j|$ and $Z_l=\sum_{j\in\mathbb Z_l}\omega_l^j|j\rangle\langle j|$ with $\omega_l=e^{i\frac{2\pi}l}$ and $\mathbb Z_l$ being the ring of addition modular $l$ which satisfies $Z_lX_l=\omega_l X_lZ_l$ and $X_l^l=Z_l^l=I$. It is easy to prove that the group $\{\{X_p,Z_p\}\otimes\{X_r,Z_r\}\}$ forms an error basis of a quqit.  Then the mixed-alphabet system, $n_1$ qupits and $n_2$ qupits, can be regarded as a composite system of an $n$-qupit and an $n_1$-qurit subsystems, with a nice error basis given by
\begin{equation}
\{\mE_p\otimes\mE_r:=X_p^{\bs}Z_p^{\bt}\otimes X_r^{\bs'}Z_r^{\bt'}\big|{\bs,\bt} \in \mathbb{Z}_{p}^{\otimes n},{\bs',\bt'} \in \mathbb{Z}_{r}^{\otimes n_1} \}.
\end{equation}
Thus any less than $d$-bit error can be regarded as two errors on two subsystems respectively, i.e.,
\begin{equation}
|\mE|=|\mE_p\cup\mE_r|=|\widehat{\bs}\cup\widehat{\bt}\cup\widehat{\bs'}\cup\widehat{\bt'}|<d,
\end{equation}
where $\widehat{\bs}=\{i\in n|s_i\neq 0\}$ is the
support of vector $\bs\in\mathbb Z_p^{\otimes n}$ and $|C|$ indicates the number
of elements in $C\subseteq n$.

For the $p$-level subsystem, consider a $\mathbb Z_p$-weighted graph $G_p=(V,\Gamma_p)$ \cite{nb3} composed of a set
$V$ of $n$ vertices and a set of weighted edges specified by the
adjacency matrix $\Gamma_p$ which is an
$n\times n$ matrix with zero diagonal entries and the matrix element
$\Gamma_{ab}\in\mathbb Z_p$ indicating the weight of the edge
connecting vertices $a$ and $b$. The graph state on $G_p$ reads $|\Gamma_p\rangle=\prod_{a,b\in V}\mathcal ({\mathcal U}_{ab})^{\Gamma_{ab}}|\theta_0\rangle^V$,
where $|\theta_0\rangle^V=(\frac1{\sqrt p}\sum_{j\in \mathbb Z_p}|j\rangle)^{\otimes n}$ is the joint $+1$ eigenstate of all $\{X_p^{\bs}|\bs\in\mathbb Z_p^{\otimes n}\}$ and $\mathcal U_{ab}=\sum_{i,j\in\mathbb Z_p}\omega_p^{ij}|i\rangle\langle i|_a\otimes |j\rangle\langle j|_b$ is the non-binary controlled phase gate between qupits $a$ and $b$. Thus $|\Gamma_p\rangle$ is the joint $+1$ eigenstate of a stabilizer group $\{g_p^\bs:=X_p^{\bs}Z_p^{\bs\cdot\Gamma_p}\big|\bs\in\mathbb{Z}_{p}^{\otimes n}\}$. Given a $\mathbb Z_r$-weighted graph $G_r=(V_1,\Gamma_r)$ for the $r$-level subsystem with $V_1\subset V$ indicating the first $n_1$ vertices of $V$ as well, then
\begin{equation}
\{Z_p^{\bc}|\Gamma_p\rangle\otimes Z_r^{\bc'}|\Gamma_r\rangle\Big|\bc\in
\mathbb Z_p^{\otimes n},\bc'\in\mathbb Z_r^{\otimes n_1}\}
\end{equation}
defines a basis of the mixed-alphabet system. Since graph states own a good feature that any bit shift error can be replaced by a phases shift error, i.e., $X^{\bs}Z^{\bt}|\Gamma\rangle\propto Z^{\bt-\bs\cdot\Gamma}|\Gamma\rangle$, it allows us to introduce a {\it composite coding clique} for the mixed-alphabet system in below.

{\bf Definition} Given graphs $G_p=(V,\Gamma_p)$ and $G_r=(V_1,\Gamma_r)$ for an $n$-qupit and an $n_1$-qurit subsystems respectively, we
define the {\it d-uncoverable set} as
\begin{equation}
\begin{split}
&\mathbb D_d=\mathbb Z_p^{\otimes n}\otimes Z_r^{\otimes n_1}-\{({\bt}-{\bs}\cdot\Gamma_p)\otimes({\bt'}-{\bs'}\cdot\Gamma_r)\big| \\
&0<|\widehat{\bs}\cup\widehat{\bt}\cup\widehat{\bs'}\cup\widehat{\bt'}|<d\}
\end{split}
\end{equation}
and the {\it d-purity set} as
\begin{equation}
\begin{split}
&\mathbb S_d=\{{\bs}\otimes{\bs'}\in \mathbb Z_p^V\otimes Z_r^{V_1}\Big|
|\widehat{\bs}\cup\widehat{\bs\cdot\Gamma_p}\cup\widehat{\bs'}\cup\widehat{\bs'\cdot\Gamma_r}|<d\}.
\end{split}
\end{equation}
A composite coding clique $\mathbb{C}^K_d$ is a collection of $K$ different vectors $\{\bc_i\otimes\bc'_i|i=1,\cdots,K\}$ in $\mathbb Z_p^{\otimes n}\otimes Z_r^{\otimes n_1}$ that satisfy:
\begin{itemize}
\item[(i)]  ${\bf 0} \in\mathbb{C}^{K}_d$;
\item[(ii)]  $\omega_p^{{\bs}\cdot{\bc}}\omega_r^{{\bs'}\cdot{\bc'}}=1$ for all ${\bs}\otimes{\bs'}\in\mathbb S_d$ and
 $\bc\otimes\bc'\in \mathbb C^K_d$;
\item[(iii)]  $(\bc_i-\bc_j)\otimes(\bc'_i\underline{\underline{}}-\bc'_j)\in \mathbb D_d$ for all $\bc_i\otimes\bc'_i, \bc_j\otimes\bc'_j\in \mathbb C^K_d$.
\end{itemize}

With this definition we have the following.

{\bf Theorem 1} Given a composite coding clique $\mathbb{C}^K_d$ on two graphs $G_p=(V,\Gamma_p)$ and $G_r=(V_1,\Gamma_r)$, the subspace spanned by the basis
\begin{equation}\label{code}
\{Z_p^{\bc}|\Gamma_p\rangle\otimes Z_r^{\bc'}|\Gamma_r\rangle\Big|\bc\otimes\bc'\in\mathbb C_d^K\}
\end{equation} defines a mixed-alphabet code $((n,K,d))_{q^{n_1}p^{n_2}}$ with $q=rp$.

\begin{proof}
We need to prove that for any error that $0<|\mE|<d$, the encoding space satisfies the Knill-Laflamme condition $\langle i|\mE|j\rangle=f(\mE)\delta_{ij}$. Firstly if the error is accidentally proportional to a stabilizer of state $|\Gamma_p\rangle\otimes|\Gamma_r\rangle$, i.e., $\mE=f(\mE)\cdot g_p^{\bs}\otimes g_r^{\bs'}$ with $f(\mE)$ being phase factor, we have
\begin{equation}
\begin{split}
&\langle i|\mE|j\rangle\\
=&f(\mE)\langle\Gamma_p|Z_p^{-\bc_i}g_p^{\bs}Z_p^{\bc_j}|\Gamma_p\rangle
\langle\Gamma_r|Z_r^{-\bc'_i}g_r^{\bs'} Z_r^{\bc'_j}|\Gamma_r\rangle\\
=&f(\mE)\omega_p^{\bs\cdot \bc_i}\omega_r^{\bs'\cdot \bc'_i}\langle\Gamma_p|Z_p^{\bc_j-\bc_i}|\Gamma_p\rangle
\langle\Gamma_r|Z_r^{\bc'_j-\bc'_i}\Gamma_r\rangle\\
=&f(\mE)\delta_{ij},
\end{split}
\end{equation}
where we have used the fact $\bs\otimes \bs^\prime\in \mathbb S_d$. Secondly if the error $\mE\propto X_p^{\bs}Z_p^{\bt}\otimes X_r^{\bs'}Z_r^{\bt'}$ is not a stabilizer of state $|\Gamma_p\rangle\otimes|\Gamma_r\rangle$, then
\begin{equation}
\begin{split}
&\langle\Gamma_p|Z_p^{-\bc_i}(X_p^{\bs}Z_p^{\bt}) Z_p^{\bc_j}|\Gamma_p\rangle\langle\Gamma_r|Z_r^{-\bc'_i}(X_r^{\bs}Z_r^{\bt}) Z_r^{\bc'_j} |\Gamma_r\rangle\\
\propto&\langle\Gamma_p|Z_p^{\bc_j-\bc_i}Z_p^{\bt-\bs\cdot\Gamma_p}|\Gamma_p\rangle\langle\Gamma_r|Z_r^{\bc'_j-\bc'_i}Z_r^{\bt'-\bs'\cdot\Gamma_r}|\Gamma_r\rangle\\
\end{split}
\end{equation}
which vanishes because condition (iii) of $\mathbb{C}^K_d$  makes at least one of $\bc_j-\bc_i+\bt-\bs\cdot\Gamma_p\neq0$ and $\bc'_j-\bc'_i+\bt'-\bs'\cdot\Gamma_r\neq0$ holds. Thus we have proved that the encoding space as in Eq.(\ref{code}) is an $((n,K,d))_{q^{n_1}p^{n_2}}$ code.
\end{proof}

In the case of coprime $p$ and $q$ ($p<q$), the method of composite coding clique does not work since the mixed-alphabet system can not be divided into some subsystems. Here we introduce an ancillary system with all the $n$ particles being of $q$ dimensions. Denote the projector of a quqit onto a qupit by $P_i$ which satisfies $P_i^\dagger=P_i$, $P_i^2=P_i$ and $Tr(P_i)=p$.  Thus the projector that projects the ancillary system onto the mixed-alphabet system reads $\mathbb P=I_1\otimes\cdots\otimes I_{n_1}\otimes P_{n_1+1}\otimes\cdots\otimes P_n$. Then we have a theorem in below.

{\bf Theorem 2} Given an $n$-quqit ancillary system, a $q^{n_1}p^{n_2}$ mixed-alphabet system and the corresponding projector $\mathbb P$, for any error on the mixed-alphabet system that $|\mE|<d$, if a $K$-dimensional encoding space of the ancillary system with basis $\{|l\rangle|l=1,\cdots,K\}$ can correct the corresponding error $\mathbb P^\dagger\mE\mathbb P$, then the subspace spanned by
\begin{equation}
\{|l'\rangle:=\frac{\mathbb P}{\sqrt{\langle l|\mathbb P|l\rangle}}|l\rangle\big|l=1,\cdots,K\}
\end{equation}
defines a mixed-alphabet code $((n,K,d))_{q^{n_1}p^{n_2}}$.

\begin{proof}
Firstly for any error $|\mE|<d$, we have
\begin{equation}
\langle i'|\mE|j'\rangle=\frac{1}{\langle i|\mathbb P|j\rangle}\langle i|\mathbb P^\dagger\mE\mathbb P|j\rangle=f(\mE)\delta_{ij}.
\end{equation}
Secondly any two basis $|i'\rangle$ and $|j'\rangle$ satisfy that
\begin{equation}
\langle i'|j'\rangle=\frac{1}{\langle i|\mathbb P|j\rangle}\langle i|\mathbb P^\dagger\mathbb P|j\rangle=0, \quad (i\neq j)
\end{equation}
which means the dimension of the encoding space of the mixed-alphabet system is still $K$. Thus this is a $((n,K,d))_{q^{n_1}p^{n_2}}$ code.
\end{proof}

Among all kinds of quantum bounds, the Singleton bound (qSB) and the Hamming bound (qHB) \cite{gottesman0} are two most important ones. Comparatively the qSB is stronger for short codes and weaker for long codes than the qHB. Since the qHB can easily be generalized to the case of mixed alphabets, here we focus on the generalization of the qSB. Consider a mixed-alphabet code with parameters $((n,K,d))_{p_1p_2\cdots p_n}$ where $p_i$ indicates the dimensions of the $i$th particle. Considering an arbitrary partition of $V=A\cup B\cup C$ into 3 parts with $|A|=|B|=d-1$ and $|C|=n-2(d-1)$.
By introducing three reduced projectors $P_A=Tr_{BC}P$, $P_B=Tr_{AC}P$ and $P_{BC}=Tr_AP$ with $P$ being the projector of the encoding subspace. For part $A$ we have
\begin{equation}
\begin{split}
TrP_A^2=&\sum_{\alpha\subseteq A}\frac{|Tr(\varepsilon_\alpha P)|^2}{K_A}=\sum_{\alpha\subseteq A}\frac{Tr(\varepsilon_\alpha P\varepsilon^{\dagger}_\alpha P)}{K_A}\\
=&KTrP^2_{BC}\geq\frac{K}{K_C}TrP^2_B,
\end{split}
\end{equation}
where the first equality is due to the expansion $P_A\propto\sum_{\alpha\subseteq A}Tr(P\varepsilon_\alpha)\varepsilon^\dagger_\alpha$ since $\{\varepsilon_\alpha\}$ forms a basis of part $A$, the second equality is due to the error-correction condition that $P\varepsilon_i\varepsilon_j^\dagger P=\frac{1}{K}Tr(P\varepsilon_i\varepsilon_j^\dagger)P$, and the inequality is due to $Tr(P_{BC}-P_B/K_C)^2\geq0$.
Similarly for part $B$, $TrP_B^2\geq\frac{K}{K_C}TrP^2_A$. Thus we have

{\bf Theorem 3} For a QECC $((n,K,d))_{p_1p_2\cdots p_n}$ over mixed alphabets it holds
\begin{equation}\label{qSB}
K\leq \min\left\{\prod_{i\in C}p_i\bigg|C\subset V, |C|=n-2(d-1)\right\}.
\end{equation}

{\it Example 1} Given two 2-weighted loop graphs $L_6$ and $L'_6$ with the corresponding vertices paired up as shown in Fig.1(A), the best composite coding clique we find for $d=3$ is a group containing 4 generators that reads $\{100100\otimes001101,$ $010010\otimes001011,$ $001101\otimes101001,$ $000110\otimes110000\}$, which means $\{Z^{14}Z^{3'4'6'},$ $Z^{25}Z^{3'5'6'},$ $Z^{346}Z^{1'3'6'},$ $Z^{45}Z^{1'2'}\}$. Then the subspace spanned by basis $\{Z^{\bc}|L_6\rangle\otimes Z^{\bc'}|L'_6\rangle\Big|\bc\otimes\bc'\in\mathbb C_3^{16}\}$ forms the optimal $((6,16,3))_4$ code. The standard stabilizer code $[[6,2,3]]_4$ has been constructed in \cite{nb6} and \cite{nb8}. While our $((6,16,3))_4$ code gives another stabilizer construction whose stabilizer has 8 generators with addition modular 2 that taks the following form.
\begin{equation}
\left[\begin{array}{c}
XZZXZZ\otimes XZIIIZ\cr
ZXZZXZ\otimes ZXZIII\cr
ZZXZZX\otimes IIIIII\cr
IIIIII\otimes ZZXZZX\cr
XZIIIZ\otimes IIZXZI\cr
ZXZIII\otimes IIIZXZ\cr
IZXZII\otimes YYZIIZ\cr
YXYZIZ\otimes IZXZII
\end{array}\right].
\end{equation}

{\it Example 2} Given two 2-weighted loop graph $L_6$ and $L'_5$ with the corresponding vertices paired up as shown in Fig.1(B), we find a $((6,8,3))_{4^52^1}$ code whose composite coding clique contains 3 generators as $\{Z^1Z^{1'2'3'4'5'},$ $Z^{2345}Z^{1'},$ $Z^{56}Z^{2'4'}\}$. Similarly via loop graphs $L_6$ and $L'_4$ paired up as Fig.1(C), we find a $((6,4,3))_{4^42^2}$ code with the composite coding clique containing 2 generators as $\{Z^{1235}Z^{2'4'},$ $Z^{2346}Z^{1'3'}\}$. Inequality (\ref{qSB}) shows that these two codes are both optimal since they saturate the quantum Singleton bound.
\begin{figure}
\begin{center}
\includegraphics[width=3.0in]{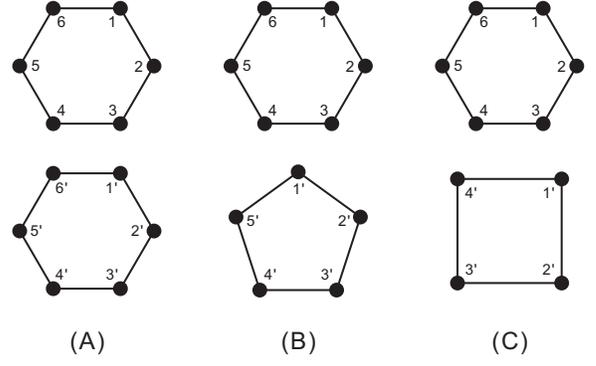}
\caption{(A) Graph of the $((6,16,3))_4$ code with corresponding vertices paired up; (B) Graph of the $((6,8,3))_{4^52^1}$ code with corresponding vertices paired up; (C) Graph of the $((6,4,3))_{4^42^2}$ code with corresponding vertices paired up.}
\end{center}
\end{figure}

{\it Example 3} For $d=2$, via two 2-weighted loop graphs paired up as Fig.2(A), the optimal $((3,4,2))_4$ code can be constructed by 2 generators $\{Z^1Z^{2'},Z^2Z^{3'}\}$. Similarly if three loop graphs are combined as shown in Fig.2(B), we can construct the optimal $((3,8,2))_8$ code with the composite coding clique containing 3 generators as $\{Z^1Z^{2'},$ $Z^2Z^{3''},$ $Z^{3'}Z^{2''}\}$. It is known that the direct product of the encoding space of two codes $((n,K,d))_p$ and $((n,K',d))_q$ constructs an $((n,KK',d))_{pq}$ code \cite{nb5}. Thus all  $((3,2^n,2))_{2^n}$ codes can be constructed via $((3,4,2))_4$ and $((3,8,2))_8$. Since all $((3,p,2))_p$ codes for odd $p$ are known, the optimal $((3,p,2))_p$ codes with $p=4k$ for any integer $k$ can be constructed.
\begin{figure}
\begin{center}
\includegraphics[width=2.8in]{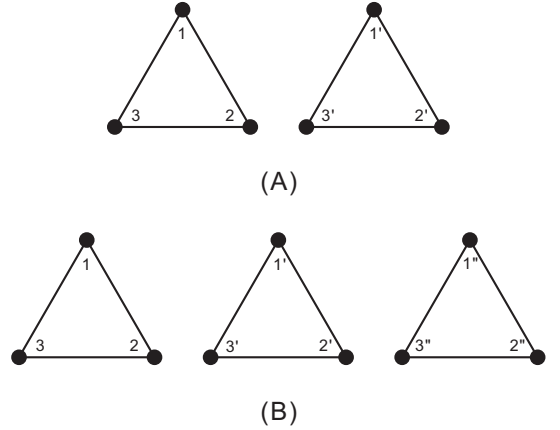}
\caption{(A) Graph of the $((3,4,2))_4$ code with corresponding vertices paired up; (B) Graph of the $((3,8,2))_8$ code with corresponding vertices combined.}
\end{center}
\end{figure}

{\it Example 4} Method of stabilizer pasting \cite{Past1,Past2} can be used to construct longer mixed-alphabet QECCs. Pasting together the $((3,4,2))_4$ code constructed above and a trivial $((2,1,2))_2$ code can construct a $((5,16,2))_{4^32^2}$ code with the stabilizer taking the following form
\begin{equation}
\left[\begin{array}{c}
Z^1Z^2X^3X^4Z^5\otimes I^{1'}I^{2'}I^{3'}\cr
I^1I^2I^3I^4I^5\otimes X^{1'}Z^{2'}Z^{3'}\cr
X^1Z^2Z^3Z^4X^5\otimes Z^{1'}X^{2'}Z^{3'}\cr
Z^1X^2Z^3I^4I^5\otimes Z^{1'}Z^{2'}X^{3'}
\end{array}\right].
\end{equation}
Pasting the $((3,4,2))_4$ code with $n$ copies of the trivial $((2,1,2))_4$ code can construct the $((2n+3,4^{2n+1},2))_4$ code.
Similarly the $((2n+3,8^{2n+1},2))_8$ code can be constructed via pasting the $((3,8,2))_8$ code with $n$ copies of the trivial $((2,1,2))_8$ code. Since all the $((2n+3,p^{2n+1},2))_p$ codes for odd $p$ are known, the $((2n+3,p^{2n+1},2))_p$ codes with $p=4k$ for any integer $k$ can be constructed.

{\it Example 5} To construct the $((5,K,2))_{3^42^1}$ code, since 3 and 2 are coprime, a 5-qutrit ancillary system is needed. For a qutrit, $X=|1\rangle\langle0|+|2\rangle\langle1|+|0\rangle\langle2|$ and $Z=|0\rangle\langle0|+\omega|1\rangle\langle1|+\omega^2|2\rangle\langle2|$ with $\omega=e^{i\frac {2\pi}3}$. Here we choose the projector as $P_i=|0\rangle\langle0|+|1\rangle\langle1|$, then
\begin{equation}
\mathbb P=I\otimes I\otimes I\otimes I\otimes P_5=2I+(1+\omega^2)Z^5+(1+\omega)(Z^5)^2
\end{equation}
up to some phase factor. For any 1-bit error $\mE$ on the first four particles, considering that $[\mE,\mathbb P]=0$, we have $\mathbb P^\dagger\mE\mathbb P=2\mE+(1+\omega^2)Z^5\mE+(1+\omega)(Z^5)^2\mE$. To the last particle, the bit flip and phase flip errors are $X'=|1\rangle\langle0|+|0\rangle\langle1|$ and $Z'=|0\rangle\langle0|-|1\rangle\langle1|$ respectively. Thus $\mathbb P^\dagger Z'\mathbb P=(1-\omega^2)Z^5+(1-\omega)(Z^5)^2$ and $\mathbb P^\dagger X'\mathbb P=X^5+X^5Z^5+X^5(Z^5)^2+(X^5)^2+\omega^2(X^5)^2Z^5+\omega(X^5Z^5)^2$ up to some phase factors. Then the problem is reduced to that of finding a code on the 5-qutrit ancillary system which can detect all 1-bit errors as well as all $\{Z_5\mE\}$ and $\{Z_5^2\mE\}$ 2-bit errors. Given a 3-weighted loop graph $L_5$ with all edges weighted 1, we find a 9-dimension subspace for the ancillary system with the coding clique as
\begin{equation}
\begin{pmatrix}
00000, &01020, &02110, &11010, &10222, \\
12200, &20210, &21102, &22120
\end{pmatrix}
\end{equation}
that can detect such errors. Thus a $((5,9,2))_{3^42^1}$ code is constructed.
However, this is a suboptimal code according to the Singleton bound. It is because any error $\mathbb P^\dagger\mE\mathbb P$ is a liner combination of $\mE,Z^5\mE$ and $(Z^5)^2\mE$, while here we have to detect each of them since the tool of coding clique we pick for the ancillary system is based on the Pauli error basis.

In this Letter, we extend the range of quantum codes to the mix-alphabet codes which have not been studied before. And the generalized quantum Singleton bound is also given. The {\it composite coding clique} is quite powerful for both mix-alphabet and standard QECCs. Many families of codes that saturate the Singleton bound can be easily constructed by this approach. However, on the other hand, the clique searching problem is intrinsically an NP-complete problem. Thus other methods need to be involved in constructing longer codes over mixed alphabets. We have used the method of stabilizer pasting to build longer codes of distance 2. And it is still an opening question to generate it to the case of $d\geq3$. To the projection-based construction, the $((5,9,2))_{3^42^1}$ code is suboptimal since the method of coding clique is not the best choice for this projector. And the problem of finding perfect method for different projectors needs to be further explored.

This work is supported by 973 Program (Grant No. 2010CB922904) and National Research Foundation and Ministry of Education, Singapore (Grant No. WBS: R-710-000-008-271).

\end{document}